# Enhancement of Water Repellence by Hierarchical Surface Structures Integrating Micro-dome and Micro-pillar Arrays with Nanoporous Coatings

*Soochan Chung[†], Kristyn Kadala[†], and Hayden Taylor[*]*

S. Chung, K. Kadala, and Prof. H. Taylor

Department of Mechanical Engineering, University of California, 6159 Etcheverry Hall, Berkeley, CA 94720, USA, and

Berkeley Education Alliance for Research in Singapore (BEARS), CREATE Tower, 1 Create Way #11-01, Singapore 138602

E-mail: hkt@berkeley.edu

[†] These authors contributed equally

**Abstract**

Superhydrophobic surfaces with multi-scale topographies offer exceptionally high apparent water contact angles and low contact angle hysteresis by virtue of the small liquid–solid contact fractions they enable. Natural water-repellent surfaces such as lotus leaves often feature dome-shaped micro-scale protrusions, whose lack of sharp edges also facilitates smooth droplet shedding without pinning. Engineered hydrophobic surfaces, however, have not yet fully exploited the merits of protrusions with controlled curvature. In this work, thermal re-flow of photoresist patterns followed by elastomeric casting was used to fabricate arrays of micro-domes with sizes 20–50 μm. These microstructures were coated with a nanoporous zinc oxide film and fluorosilanized to produce hierarchical surface topographies with static water contact angles up to 169.7±0.4° and contact angle hysteresis as low as 14.7±1.3°. Performance of the micro-dome arrays significantly exceeded that of arrays of sharp-edged square pillars and flat surfaces coated with the same nanoporous film. The highest performance came from the smallest micro-domes (20 μm) and closest spacings (10 μm) investigated. For larger features, contact angles reduced and hysteresis increased — unexpected trends not explained by contact fraction alone. This simple fabrication technique could be adapted to manufacture large surfaces for droplet shedding, including in heat transfer applications.

**Introduction**

As the use of air conditioning rises globally, technologies are desired to make cooling and dehumidification more energy-efficient.[1] One particular challenge is the accumulation of condensed atmospheric water on heat exchanger surfaces, which can impede heat transfer and permit biological contamination.[2] Scalable, robust technologies are desired to enhance condensate shedding and keep cooling coil surfaces as dry as possible. One promising route is to render surfaces superhydrophobic to promote dropwise condensation and shedding, and to this end many chemical and morphological modifications have been investigated.[3–5] Inspiration has been sought from nature,[6] in which the multi-scale surface topographies of certain leaves, such as that of the lotus, have been shown to be promote droplet coalescence and shedding by greatly reducing liquid–solid contact and minimizing droplet pinning.[7–9] Synthetic repellent surfaces generally involve sharp-edged micro- and nano-structures,[10] which can be simple to manufacture but do not deliver the same droplet-shedding performance as the smoother, dome-shaped microstructures found in nature.[11–13] Engineered dome- or dimple-shaped[14,15] and hierarchical, multi-scale[16] structures are therefore of increasing interest. Here, we show how dome-shaped microstructures can be simply fabricated and then coated with a nanoporous, fluorosilane-terminated zinc oxide film whose ultrahydrophobic properties on a flat surface we have reported previously,[17] to generate a surface with lotus leaf-like performance. The addition of a micro-dome array substantially improved hydrophobicity, as seen through increased static water contact angle and reduced contact angle hysteresis relative to both a flat surface coated with the nanoporous film and an array of sharp-edged micropillars also coated with the film. A scale dependence was also seen, with the highest performance from the smallest features in the 20–50 μm range studied.

Sometimes underlying the design of hierarchical surfaces has been the widely known Cassie–Baxter relation,[18] which models how the apparent static contact angle, $\theta^*$, of a droplet suspended on the protrusions of a rough surface increases as the fraction, $\phi$, of the projected surface area in solid–liquid contact reduces:

$$\cos \theta^* = \phi \cos \theta_0 - 1 + \phi \tag{1}$$

where $\theta_0$ is the contact angle on a smooth surface of the same material. This model has, however, been extensively challenged since shortly after its introduction, on the basis that the characteristics of the liquid–solid contact *line* — and not the contact *area* — are of central importance in determining static, advancing and receding contact angles and hence droplet-shedding effectiveness.[13,19,20] The particular geometric design[21] of surface features has been shown to influence static contact angle, sliding angle, and contact angle hysteresis (the difference between advancing and receding angles). Such dependences are seen even at a fixed contact area fraction, indicating the importance of the shape and length of the contact line.

A related concern is droplet pinning, which tends to occur at feature corners and disrupts droplet shedding even if the static contact angle is high.[13,22] Also relevant are the robustness of the droplet's state to external mechanical energy inputs (*e.g.* droplet impact[23] or hydrostatic pressure[24]), and the fact that condensing droplets may nucleate at any location on the surface and not necessarily on the tips of surface protrusions.[5,25,26] A comprehensive and robust predictive model for contact angles, droplet shedding, and robustness therefore remains elusive, and especially so for surfaces with curved topographies.

In this work, we empirically investigated the relationship between surface microstructure geometry and water droplet behavior for surfaces covered with domes and square pillars. The fabrication process for domes (Figure 1a) relies on the thermal re-flow of a single, patterned photoresist layer in which the surface tension of heated photoresist drives an energy minimization to create features with curved surfaces. A two-stage elastomeric casting process creates replicas of these microstructures, which are then sputtered with an aluminum seed layer and coated with a nanoporous ZnO film via immersion in a heated equimolar aqueous zinc nitrate/hexamine solution. This process flow allowed structures with a range of micro- and nano-scale features to be prototyped rapidly. A total of 16 micro-dome patterns were tested (*e.g.* Figure 1c), covering each possible combination of the pre-reflow feature diameters of {20, 30, 40, 50} μm with gap-to-diameter ratios of {0.5, 1.0, 1.5, 2.0}. A set of patterns was also fabricated without the thermal re-flow step, to provide square pillars with sharper corners for comparison (Figure 1b).

Optical microscopy of water droplets resting on the micropatterned surfaces indicated that they remained suspended on the tips of the structures and did not infiltrate the gaps between microfeatures. All of the microstructured surfaces tested — including pillar and dome arrays — showed higher average static water contact angles (Figure 2) than a microscopically flat surface processed in the same batch of samples and bearing only the nanoporous ZnO film. The micro-dome arrays, however, performed significantly better than the square pillars for any given nominal feature size and spacing. These general differences are compatible with existing conceptions of the role of partial liquid–solid contact.[13,18] Adding any sort of microstructure is expected to reduce both linear and areal liquid–solid contact fractions, provided that the droplet is suspended on the tips of the microstructure. Furthermore, the curvature of the domed features makes it possible for only a small region at the tip of each dome to be in contact with the liquid, whereas the entire upper surface of a flat-tipped square pillar beneath a droplet is expected to be in contact with the liquid.

There are two trends, however, that cannot be simply explained by either the contact area fraction-based Cassie–Baxter model[18] or by a model based on a linear contact fraction.[13] The first trend is a scale dependence: average contact angle reduced by about 4–5° as the nominal feature size grew from 20 to 50 μm, for a fixed gap-to-size ratio. This trend is unexpected, since no model that considers only a solid–liquid contact *fraction* can explain this dependence, and indeed previous experimental results from Hisler et al.[27] showed no such scale dependence for flat-topped pillars with widths of 4–128 µm.

A second unexpected trend is seen in the micro-domed samples: in general, as the gap-to-size ratio increased from 0.5 to 2, the average contact angle decreased by about 3–4° for a given pattern size. This trend is the opposite of what might be anticipated from existing models and prior experiments with square pillar arrays,[28] since spacing features of a particular size further apart would usually reduce both the linear and areal contact fractions and hence be expected to raise the apparent contact angle. A possible explanation is that the contact area within each individual dome increased as the domes were spaced further apart, since a greater fraction of the droplet's weight would have needed to be carried by each feature. The highest static contact angle obtained in this work, 169.7±0.4° (mean ± standard error of the mean, N = 5), occurred when the domes had the smallest diameter (20 µm) and gap (10 µm), and is slightly superior to the 164° measured from natural lotus leaves.[12] This optimal result was more than 10° higher than obtained with the 'flat' (nanostructure-only) surface.

Contact angle hysteresis was lower for both domed and pillared surfaces than for a flat surface (Figure 3). Lower hysteresis is associated with greater ease of droplet shedding,[13,29] suggesting that these microstructures may be attractive for enhancing condensate removal from a surface. The micro-domed surfaces showed consistently lower hysteresis values than square-pillared surfaces with equivalent diameters and spacings. As with the static contact angle results, smaller features and smaller gaps yielded more desirable performance in the micro-domed surfaces: the lowest hysteresis obtained was 14.7±1.3° for 20 µm-diameter domes spaced by 10 µm, whereas the hysteresis on a flat surface was far higher, at 39.0±0.4°. (For comparison, the hysteresis of the lotus leaf has been reported to be 3°.[12]) A scale-dependence was evident in the square-pillar arrays as well, with smaller pillars almost always offering lower hysteresis at any given gap-to-size ratio. Such a scale dependence was not seen, in contrast, in the experiments of Yeh[30] in which square-tipped silanized silicon pillars ranging from 3–9 µm in size were tested. The improvements in hysteresis achieved by patterning the surfaces are attributable to increased receding contact angles (Figure 3b, d); the advancing angles remained very close to that of a flat surface. The dominant role of the receding angle is consistent with, *e.g.*, Dorrer's and Rühe's experimental results for square pillars.[31]

This study suggests that micro-scale domes can usefully be added to a surface to increase static water contact angle by at least 12° and reduce contact angle hysteresis by at least 24° relative to a flat surface with a comparable nano-scale surface structure and chemistry. Best performance was obtained with the smallest (20 µm) and most closely spaced (10 µm) domes tested, suggesting that it would be useful to investigate whether further reducing dome size or spacing could increase the performance even more. The strong feature-size dependences that we have observed of both static contact angle and hysteresis are absent in much previous work, such as Hisler's[27], Yeh's[30] and Lv's.[32] That previous work was conducted on microscale pillar arrays without a secondary nanostructure, and it is possible that addition of the nanostructured film to our surfaces may, by roughening the edges of the micro-scale pillars, render the pillars' sizes more critical to the overall surface performance.

High static contact angles and low hysteresis and are widely associated with dropwise condensation and more effective droplet shedding respectively, suggesting that it may be possible to apply these geometries to enhance condensate shedding from evaporator coil surfaces during the cooling of moist air. The structures tested here involved depositing a thin aluminum layer onto an elastomeric substrate, but for heat transfer applications similar structures might be fabricated at scale in bulk aluminum alloy by, *e.g.*, coining or knurling. The demonstrated process therefore represents a step towards scalable manufacturing of practicable, water-repelling materials.

**Experimental section**

*Photolithography and thermal re-flow*

Two Mylar transparency masks were prepared: one with square arrays of square features, to produce the sharp-tipped pillars, and the other with square arrays of circular features, to produce the micro-dome arrays. Each mask contained 16 different pattern arrays sized 15 mm × 15 mm, covering every combination of the feature side-length or diameter set {20, 30, 40, 50} μm with inter-feature spacings of 0.5, 1.0, 1.5 and 2.0 times the feature size.

Two 100 mm-diameter silicon wafers were cleaned by sonication in acetone and isopropanol, rinsed with deionized (DI) water, and dehydrated at 150 °C on a hotplate for 15 min. Then, the wafers were oxygen plasma-treated at 70 W, 200 mTorr for 5 min ($O_2$ plasma system, Plasma Equipment Technical Services (PETS) Inc.). An adhesion promoter, hexamethyldisilazane (HMDS, Sigma-Aldrich), was vapor-phase coated onto the silicon wafers for 5 min.

AZ P4620 photoresist (PR) was spin-coated onto one of the wafers at 300 rpm for 19 sec, attained at a ramp rate of 50 rpm/s, followed by 850 rpm for 39 sec, attained at 100 rpm/s, with a target thickness of 20 μm. The coated wafer was soft-baked at 90 °C for 30 min. To prevent cracking of the PR, the wafer was placed in a dark room for 10 min at 30–50% relative humidity to rehydrate. This wafer was then exposed, using the square-pillar mask design, to 500 mJ/cm$^2$ i-line (365 nm) UV in a contact mask aligner (Model 200, OAI) and developed for 5 min with AZ 1400K diluted 1:3 by volume in DI water. The developed wafer was rinsed with DI water and treated with $O_2$ plasma at 60 W, 200 mTorr for 10 min ($O_2$ plasma system, Plasma Equipment Technical Services (PETS) Inc.) to descum it.

This wafer was then baked at 135 °C for 1 hour, triggering a thermal reflow driven by the PR's surface tension to obtain curved micro-dome features (Figure 1a–c). The heights of the reflowed geometries ranged between 16.5 and 25.0 µm as determined by electron microscopy (Figure 2) and stylus profilometer (Dektak 3030).

To match the heights of the micro-pillar structures to those of the reflowed domes, a different spin coating speed was used: 1000 rpm instead of 850 rpm, based on the resist manufacturer's spin curve. The other lithographic process steps were as above, but without thermal reflow step. Vertical-walled PR features resulted, and the resist height was confirmed to be 16.75 ± 0.20 µm with a surface profilometer (Dektak 3030).

*Transfer casting*

Replicas of the micro-pillar and micro-dome structures were created in polydimethylsiloxane (PDMS, Sylgard 184, Dow Corning) via two-stage casting.[33] In the first casting step, the PDMS pre-polymer and crosslinker were mixed in the ratio 5:1 and poured onto the wafer. After curing at 70 °C for 2 hours, this casting was oxygen plasma-treated (60 W, 200 mTorr, 2 min) and silanized with 1H,1H,2H,2H-perfluorooctyltrichlorosilane (Sigma-Aldrich). This casting served as the mold for a second casting step using a less rigid mixture of PDMS (pre-polymer:crosslinker :: 10:1).

*ZnO nanostructure synthesis*

The microstructured PDMS substrates were sputtered with 10 nm Cr followed by 150 nm 99.999%-pure aluminum to support the hydrothermal synthesis of porous ZnO. The growth of ZnO followed our previously reported process[17], in which the aluminum-coated PDMS was immersed in a 25 mM equimolar aqueous solution of zinc nitrate ($Zn(NO_3)_2 \cdot 6\ H_2O$, Sigma-Aldrich) and hexamine (hexamethylenetetramine, Sigma-Aldrich) at 70 °C for 90 minutes in an oven. Following this bath synthesis, samples were rinsed in DI water, dried in a jet of $N_2$, and stored at room temperature. Electron micrographs of the resulting nanostructures are shown in Figure 1d–f.

*Surface fluorosilanization*

Sample surfaces were cleaned with oxygen plasma (60 W, 200 mTorr, 2 min). To achieve surface superhydrophobicity, samples were then immediately placed in a vacuum desiccator with 100 μL of 1H,1H,2H,2H-perfluorooctyltrichlorosilane (Sigma-Aldrich), pumped down for 20 minutes and left to rest for 40 minutes. Samples were rinsed with DI water and allowed to 'cure' at room temperature in a fumehood for 24 hours.

*Contact angle measurement*

Static, advancing, and receding water contact angles were measured using a custom-built goniometer. Five droplets (7.5 μL) were sequentially deposited onto different parts of each sample surface and measured. Video images were captured of each droplet, viewed from the edge of the substrate using a Thorlabs DCC1645C camera with a 25 mm focal-length plano-convex lens. Images were analyzed in ImageJ.[34] Static contact angles were extracted using the Low-Bond Axisymmetric Drop Shape Analysis (LB-ADSA) ImageJ plug-in,[35] which fits the Young–Laplace equation to image data. The sample stage was then tilted until the droplet rolled off, and the Dropsnake ImageJ plugin[36] was used to extract advancing and receding contact angles from the video frame captured immediately prior to roll-off. Hysteresis is defined as the difference between advancing and receding angles.

**Acknowledgements**

The authors thank the staff of the U.C. Berkeley Biomolecular Nanotechnology Center (BNC) for assistance. The authors acknowledge helpful discussions with Lance Brockway and Zahra Hemmatian. This work was supported by the Singapore–Berkeley Building Efficiency and Sustainability in the Tropics (SinBerBEST) program, funded by the National Research Foundation, Prime Minister's Office,

Singapore. H.T. is an inventor on U.S. patent application 62/207,109, which relates to ZnO nanoporous structures. The authors declare no other conflicts of interest.

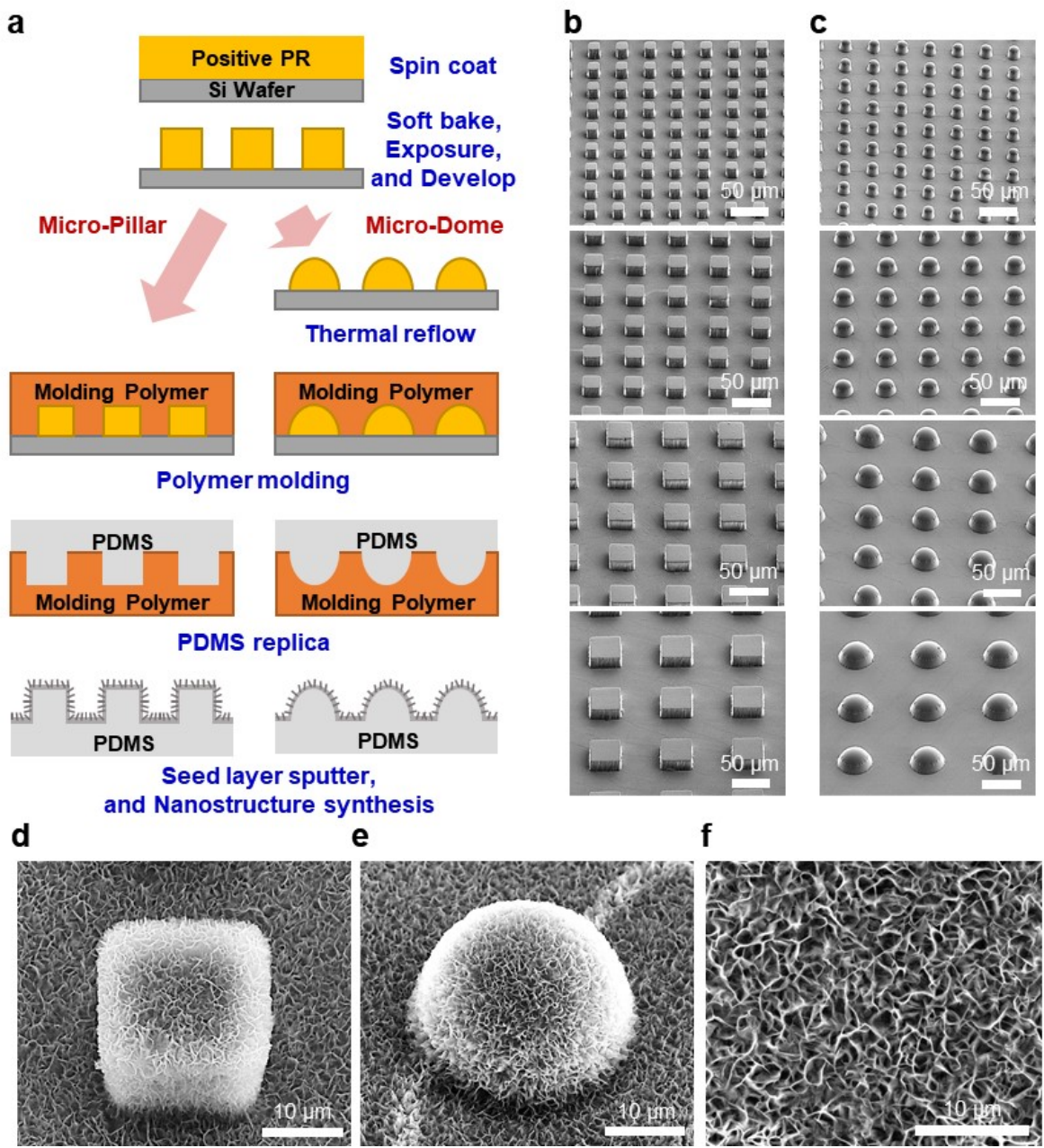

**Figure 1. Fabrication and characterization of hierarchical structures.** (a) Process flow for both dome and square pillar arrays; (b) Scanning electron microscope (SEM) images showing micro-pillar arrays with pattern sizes of (from top to bottom) 20 µm, 30 µm, 40 µm, 50 µm. Gap-to-pattern size ratio was 1 in all cases. Scale bar: 50µm; (c) SEM images showing micro-dome arrays with the same set of pattern and gap sizes as (b); (d, e) SEM images of a representative square pillar (d) and dome (e), showing the nanoscale porosity of the deposited ZnO film. Scale bar: 10µm; (f) Enlarged view of nanoporous ZnO coating. Scale bar: 10 µm.

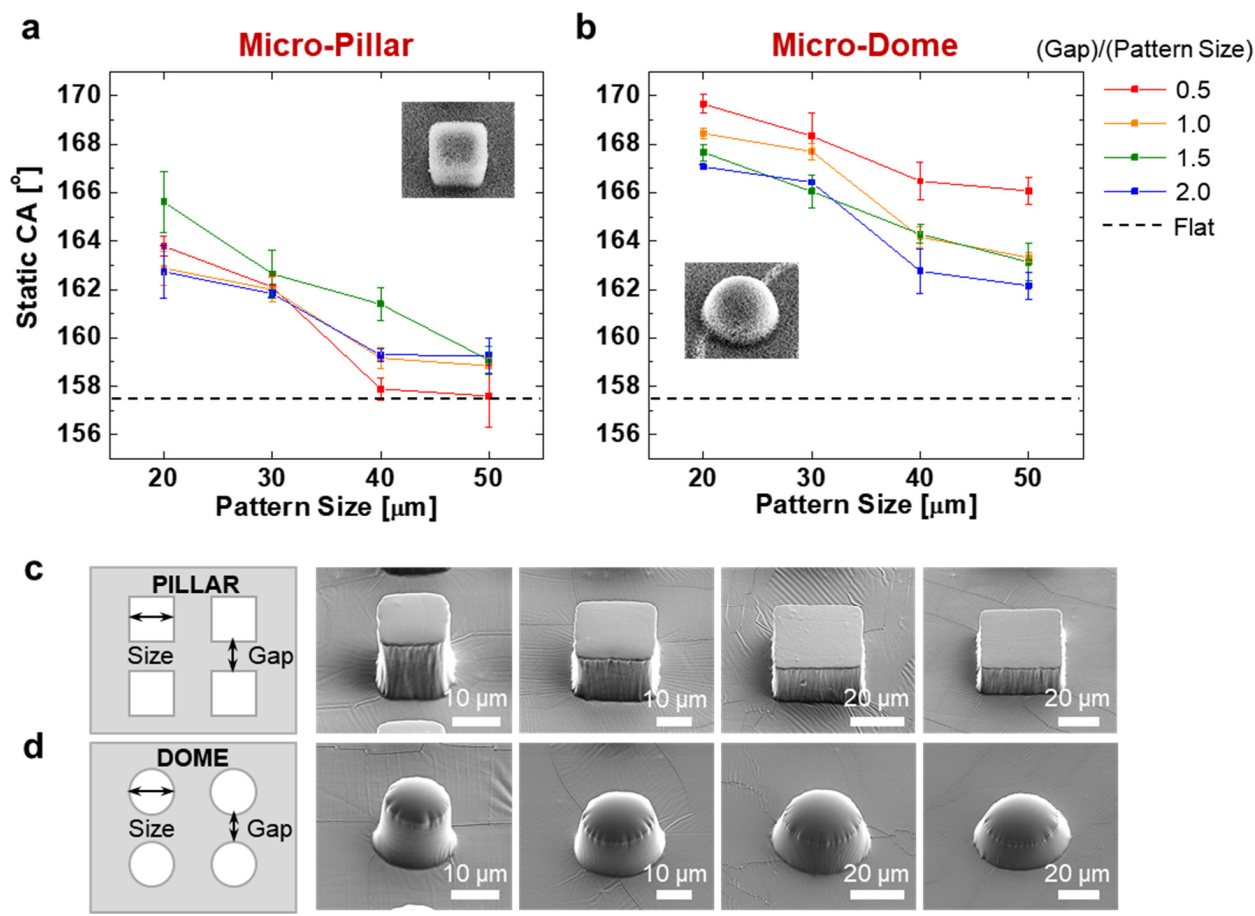

**Figure 2.** Water sessile contact angles on arrays of (a) square-tipped pillars and (b) micro-domes, both covered with a fluorosilanized ZnO nanoporous coating. Contact angles are plotted against feature diameter ('pattern size') for varying gap-to-pattern-size ratio. Error bars represent ±1 standard error of the mean; five droplets per specimen. 'Flat' denotes the contact angle on a surface with the fluorosilanized ZnO coating but no microfeatures. SEM images of representative pillars (c) and domes (d) are shown after transfer casting and aluminum sputtering but before coating with the ZnO film. The nominal sizes of the features are {20, 30, 40, 50} µm from left to right. The center heights of the domes after re-flow were {16.52, 17.57, 19.88, 25.01} µm and the apex radii of curvature were {9.02, 12.52, 18.16, 25.39} µm. The heights of the square pillars were adjusted during fabrication to match those of the re-flowed domes.

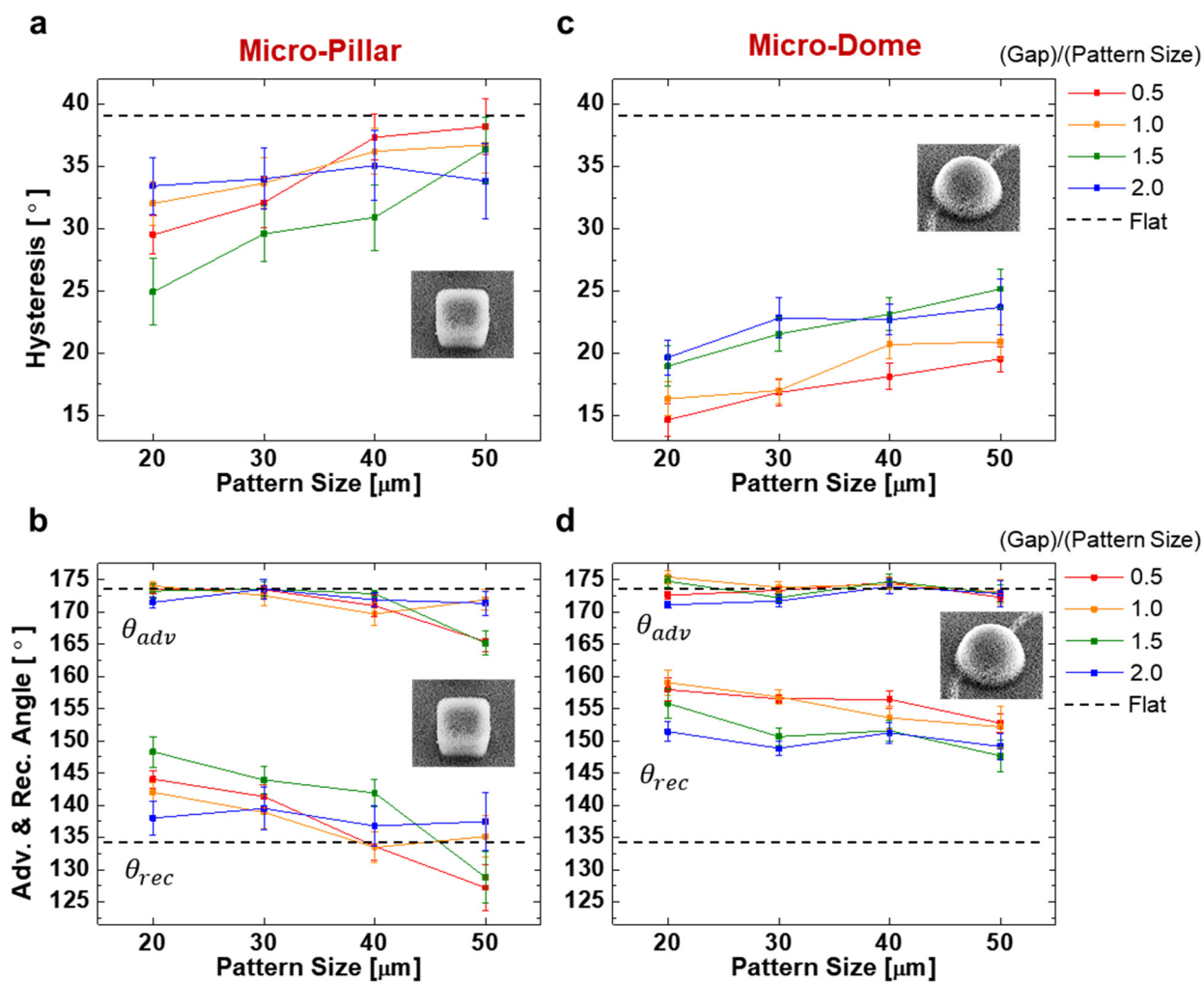

Figure 3. Contact angle hysteresis and advancing and receding contact angles for arrays of (a, b) square pillars and (c, d) domes, all covered with a fluorosilanized ZnO nanoporous coating. Results are plotted against pattern size for a range of gap-to-pattern-size ratios. Error bars represent ±1 standard error of the mean; sample size is five separate droplet sheddings per specimen. 'Flat' denotes the corresponding results on a surface with the fluorosilanized ZnO coating but no microfeatures.

**Table of Contents text:**

Arrays of 20–50 μm micro-domes were coated with nanoporous zinc oxide and fluorosilanized, producing hierarchical superhydrophobic surfaces. Water contact angles up to 169.7±0.4° and hysteresis as low as 14.7±1.3° were achieved, outperforming arrays of sharp-edged square pillars and flat surfaces coated with the nanoporous film. Smaller micro-domes with closer spacings offered the highest performance — unexpected trends that cannot be explained by purely contact fraction-based models.

**Table of Contents figure:**

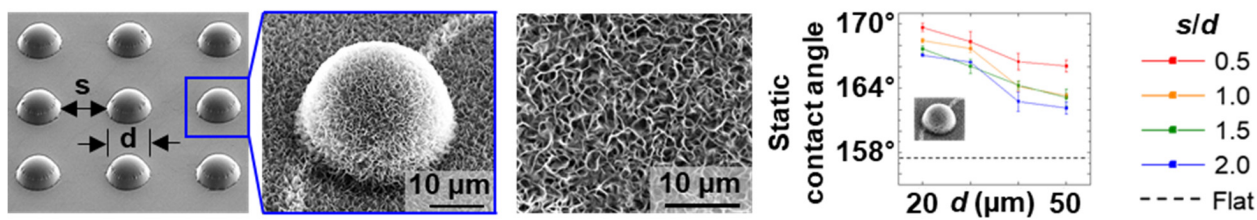

**Table of Contents keyword:**

Superhydrophobicity